



\font\bigbf=cmbx10 scaled\magstep2

\font\twelverm=cmr10 scaled 1200    \font\twelvei=cmmi10 scaled 1200
\font\twelvesy=cmsy10 scaled 1200   \font\twelveex=cmex10 scaled 1200
\font\twelvebf=cmbx10 scaled 1200   \font\twelvesl=cmsl10 scaled 1200
\font\twelvett=cmtt10 scaled 1200   \font\twelveit=cmti10 scaled 1200

\skewchar\twelvei='177   \skewchar\twelvesy='60


\def\twelvepoint{\normalbaselineskip=12.4pt
  \abovedisplayskip 12.4pt plus 3pt minus 9pt
  \belowdisplayskip 12.4pt plus 3pt minus 9pt
  \abovedisplayshortskip 0pt plus 3pt
  \belowdisplayshortskip 7.2pt plus 3pt minus 4pt
  \smallskipamount=3.6pt plus1.2pt minus1.2pt
  \medskipamount=7.2pt plus2.4pt minus2.4pt
  \bigskipamount=14.4pt plus4.8pt minus4.8pt
  \def\rm{\fam0\twelverm}          \def\it{\fam\itfam\twelveit}%
  \def\sl{\fam\slfam\twelvesl}     \def\bf{\fam\bffam\twelvebf}%
  \def\mit{\fam 1}                 \def\cal{\fam 2}%
  \def\tt{\twelvett}
  \textfont0=\twelverm   \scriptfont0=\tenrm   \scriptscriptfont0=\sevenrm
  \textfont1=\twelvei    \scriptfont1=\teni    \scriptscriptfont1=\seveni
  \textfont2=\twelvesy   \scriptfont2=\tensy   \scriptscriptfont2=\sevensy
  \textfont3=\twelveex   \scriptfont3=\twelveex
 \scriptscriptfont3=\twelveex
  \textfont\itfam=\twelveit
  \textfont\slfam=\twelvesl
  \textfont\bffam=\twelvebf \scriptfont\bffam=\tenbf
  \scriptscriptfont\bffam=\sevenbf
  \normalbaselines\rm}



\def\beginlinemode{\endmode
  \begingroup\parskip=0pt
\obeylines\def\\{\par}\def\endmode{\par\endgroup}}
\def\beginparmode{\endmode
  \begingroup \def\endmode{\par\endgroup}}
\let\endmode=\par
{\obeylines\gdef\
{}}
\def\singlespace{\baselineskip=\normalbaselineskip}
\def\oneandathirdspace{\baselineskip=\normalbaselineskip
  \multiply\baselineskip by 4 \divide\baselineskip by 3}
\def\oneandahalfspace{\baselineskip=\normalbaselineskip
  \multiply\baselineskip by 3 \divide\baselineskip by 2}
\def\doublespace{\baselineskip=
\normalbaselineskip \multiply\baselineskip by 2}

\newcount\firstpageno
\firstpageno=1
\footline={\ifnum\pageno<\firstpageno{\hfil}%
\else{\hfil\twelverm\folio\hfil}\fi}
\let\rawfootnote=\footnote              
\def\footnote#1#2{{\rm\singlespace\parindent=0pt\rawfootnote{#1}{#2}}}
\def\raggedcenter{\leftskip=4em plus 12em \rightskip=\leftskip
  \parindent=0pt \parfillskip=0pt \spaceskip=.3333em \xspaceskip=.5em
  \pretolerance=9999 \tolerance=9999
  \hyphenpenalty=9999 \exhyphenpenalty=9999 }
\def\dateline{\rightline{\ifcase\month\or
  January\or February\or March\or April\or May\or June\or
  July\or August\or September\or October\or November\or December\fi
  \space\number\year}}
\def\received{\vskip 3pt plus 0.2fill
 \centerline{\sl (Received\space\ifcase\month\or
  January\or February\or March\or April\or May\or June\or
  July\or August\or September\or October\or November\or December\fi
  \qquad, \number\year)}}


\hsize=6.5truein
\vsize=8.9truein
\voffset=0.0truein
\parskip=\medskipamount
\twelvepoint            
\oneandathirdspace           
\overfullrule=0pt       



\def\title                      
  {\null\vskip 3pt plus 0.2fill
   \beginlinemode \doublespace \raggedcenter \bigbf}

\def\author                     
  {\vskip 3pt plus 0.2fill \beginlinemode
   \singlespace \raggedcenter}

\def\affil                      
  {\vskip 3pt plus 0.1fill \beginlinemode
   \oneandahalfspace \raggedcenter \sl}

\def\abstract                   
  {\vskip 3pt plus 0.3fill \beginparmode
   \oneandathirdspace\narrower}

\def\endtitlepage               
  {\endpage                     
   \body}

\def\body                       
  {\beginparmode}               

\def\subhead#1{                 
  \vskip 0.25truein             
  \noindent{{\it {#1}} \par}
   \nobreak\vskip 0.15truein\nobreak}

\def\refto#1{[#1]}           

\def\references                 
  {\subhead{\bf References}         
   \beginparmode
   \frenchspacing \parindent=0pt \leftskip=1truecm
   \oneandathirdspace\parskip=8pt plus 3pt
 \everypar{\hangindent=\parindent}}

\gdef\refis#1{\indent\hbox to 0pt{\hss#1.~}}    

\gdef\journal#1, #2, #3, 1#4#5#6{               
    {\sl #1~}{\bf #2}, #3 (1#4#5#6)}           

\def\refstylenp{                
  \gdef\refto##1{ [##1]}                                
  \gdef\refis##1{\indent\hbox to 0pt{\hss##1)~}}        
  \gdef\journal##1, ##2, ##3, ##4 {                     
     {\sl ##1~}{\bf ##2~}(##3) ##4 }}

\def\refstyleprnp{              
  \gdef\refto##1{ [##1]}                                
  \gdef\refis##1{\indent\hbox to 0pt{\hss##1)~}}        
  \gdef\journal##1, ##2, ##3, 1##4##5##6{               
    {\sl ##1~}{\bf ##2~}(1##4##5##6) ##3}}

\def\prl{\journal Phys. Rev. Lett., }

\def\jmp{\journal J. Math. Phys., }

\def\ann{\journal Ann. Phys., }

\def\grg{\journal Gen. Rel. Grav., }

\def\endreferences{\body}

\def\figurecaptions             
  { \beginparmode
   \subhead{Figure Captions}
}

\def\endpage                    
  {\vfill\eject}

\def\endpaper                   
  {\endmode\vfill\supereject}


\def\ref#1{Ref. #1}                     
\def\Ref#1{Ref. #1}                     

\def\frac#1#2{{\textstyle{#1 \over #2}}}

\def\sla{\raise.15ex\hbox{$/$}\kern-.57em}
\def\leaderfill{\leaders\hbox to 1em{\hss.\hss}\hfill}
\def\twiddle{\lower.9ex\rlap{$\kern-.1em\scriptstyle\sim$}}
\def\bigtwiddle{\lower1.ex\rlap{$\sim$}}
\def\gtwid{
\mathrel{\raise.3ex\hbox{$>$\kern-.75em\lower1ex\hbox{$\sim$}}}}
\def\ltwid{\mathrel{\raise.3ex\hbox
{$<$\kern-.75em\lower1ex\hbox{$\sim$}}}}
\def\square{\kern1pt\vbox{\hrule height 1.2pt\hbox
{\vrule width 1.2pt\hskip 3pt
   \vbox{\vskip 6pt}\hskip 3pt\vrule width 0.6pt}
\hrule height 0.6pt}\kern1pt}

\def\m@th{\mathsurround=0pt }
\def\leftrightarrowfill{$\m@th \mathord\leftarrow \mkern-6mu
 \cleaders\hbox{$\mkern-2mu \mathord- \mkern-2mu$}\hfill
 \mkern-6mu \mathord\rightarrow$}
\def\overleftrightarrow#1{\vbox{\ialign{##\crcr
     \leftrightarrowfill\crcr\noalign{\kern-1pt\nointerlineskip}
     $\hfil\displaystyle{#1}\hfil$\crcr}}}


\font\titlefont=cmr10 scaled\magstep3

\def\martinstyletitle                      
  {\null\vskip 3pt plus 0.2fill
   \beginlinemode \doublespace \raggedcenter \titlefont}

\font\twelvesc=cmcsc10 scaled 1200

\def\author                     
  {\vskip 3pt plus 0.2fill \beginlinemode
   \singlespace \raggedcenter\twelvesc}


\def\heading                            
  {\vskip 0.5truein plus 0.1truein      
\endheading
   \beginparmode \def\\{\par} \parskip=0pt \singlespace \raggedcenter}

\def\endheading
  {\par\nobreak\vskip 0.25truein\nobreak\beginparmode}

\def\subheading                         
  {\vskip 0.25truein plus 0.1truein
   \beginlinemode \singlespace \parskip=0pt \def\\{\par}\raggedcenter}

\def\tag#1$${\eqno(#1)$$}

\def\align#1$${\eqalign{#1}$$}

\def\aligntag#1$${\gdef\tag##1\\{&(##1)\cr}\eqalignno{#1\\}$$
  \gdef\tag##1$${\eqno(##1)$$}}

\def\endaligntag{}

\def\overset #1\to#2{{\mathop{#2}\limits^{#1}}}
\def\underset#1\to#2{{\let\next=#1\mathpalette\undersetpalette#2}}
\def\undersetpalette#1#2{\vtop{\baselineskip0pt
\ialign{$\mathsurround=0pt #1\hfil##\hfil$\crcr#2\crcr\next\crcr}}}


\def\ref#1{Ref.~#1}                     
\def\Ref#1{Ref.~#1}                     
\def\[#1]{[\cite{#1}]}
\def\cite#1{{#1}}
\def\(#1){(\call{#1})}
\def\call#1{{#1}}
\def\taghead#1{}
\def\frac#1#2{{#1 \over #2}}

\def\12{{1\over2}}

\def\sla{\raise.15ex\hbox{$/$}\kern-.57em}
\def\leaderfill{\leaders\hbox to 1em{\hss.\hss}\hfill}
\def\twiddle{\lower.9ex\rlap{$\kern-.1em\scriptstyle\sim$}}
\def\bigtwiddle{\lower1.ex\rlap{$\sim$}}
\def\gtwid{\mathrel{\raise.3ex\hbox{$>$
\kern-.75em\lower1ex\hbox{$\sim$}}}}
\def\ltwid{\mathrel{\raise.3ex\hbox{$<$
\kern-.75em\lower1ex\hbox{$\sim$}}}}
\def\square{\kern1pt\vbox{\hrule height 1.2pt\hbox
{\vrule width 1.2pt\hskip 3pt
   \vbox{\vskip 6pt}\hskip 3pt\vrule width 0.6pt}
\hrule height 0.6pt}\kern1pt}
\def\tdot#1{\mathord{\mathop{#1}\limits^{\kern2pt\ldots}}}

\def\pmb#1{\setbox0=\hbox{#1}%
  \kern-.025em\copy0\kern-\wd0
  \kern  .05em\copy0\kern-\wd0
  \kern-.025em\raise.0433em\box0 }

\catcode`@=11
\newcount\tagnumber\tagnumber=0

\immediate\newwrite\eqnfile
\newif\if@qnfile\@qnfilefalse
\def\write@qn#1{}
\def\writenew@qn#1{}
\def\w@rnwrite#1{\write@qn{#1}\message{#1}}
\def\@rrwrite#1{\write@qn{#1}\errmessage{#1}}

\def\taghead#1{\gdef\t@ghead{#1}\global\tagnumber=0}
\def\t@ghead{}

\expandafter\def\csname @qnnum-3\endcsname
  {{\t@ghead\advance\tagnumber by -3\relax\number\tagnumber}}
\expandafter\def\csname @qnnum-2\endcsname
  {{\t@ghead\advance\tagnumber by -2\relax\number\tagnumber}}
\expandafter\def\csname @qnnum-1\endcsname
  {{\t@ghead\advance\tagnumber by -1\relax\number\tagnumber}}
\expandafter\def\csname @qnnum0\endcsname
  {\t@ghead\number\tagnumber}
\expandafter\def\csname @qnnum+1\endcsname
  {{\t@ghead\advance\tagnumber by 1\relax\number\tagnumber}}
\expandafter\def\csname @qnnum+2\endcsname
  {{\t@ghead\advance\tagnumber by 2\relax\number\tagnumber}}
\expandafter\def\csname @qnnum+3\endcsname
  {{\t@ghead\advance\tagnumber by 3\relax\number\tagnumber}}

\def\equationfile{%
  \@qnfiletrue\immediate\openout\eqnfile=\jobname.eqn%
  \def\write@qn##1{\if@qnfile\immediate\write\eqnfile{##1}\fi}
  \def\writenew@qn##1{\if@qnfile\immediate\write\eqnfile
    {\noexpand\tag{##1} = (\t@ghead\number\tagnumber)}\fi}
}

\def\callall#1{\xdef#1##1{#1{\noexpand\call{##1}}}}
\def\call#1{\each@rg\callr@nge{#1}}

\def\each@rg#1#2{{\let\thecsname=#1\expandafter\first@rg#2,\end,}}
\def\first@rg#1,{\thecsname{#1}\apply@rg}
\def\apply@rg#1,{\ifx\end#1\let\next=\relax%
\else,\thecsname{#1}\let\next=\apply@rg\fi\next}

\def\callr@nge#1{\calldor@nge#1-\end-}
\def\callr@ngeat#1\end-{#1}
\def\calldor@nge#1-#2-{\ifx\end#2\@qneatspace#1 %
  \else\calll@@p{#1}{#2}\callr@ngeat\fi}
\def\calll@@p#1#2{\ifnum#1>#2{\@rrwrite
{Equation range #1-#2\space is bad.}
\errhelp{If you call a series of equations by the notation M-N, then M and
N must be integers, and N must be greater than or equal to M.}}\else %
{\count0=#1\count1=
#2\advance\count1 by1\relax\expandafter\@qncall\the\count0,%
  \loop\advance\count0 by1\relax%
    \ifnum\count0<\count1,\expandafter\@qncall\the\count0,%
  \repeat}\fi}

\def\@qneatspace#1#2 {\@qncall#1#2,}
\def\@qncall#1,{\ifunc@lled{#1}{\def\next{#1}\ifx\next\empty\else
  \w@rnwrite{Equation number \noexpand\(>>#1<<)
has not been defined yet.}
  >>#1<<\fi}\else\csname @qnnum#1\endcsname\fi}

\let\eqnono=\eqno
\def\eqno(#1){\tag#1}
\def\tag#1$${\eqnono(\displayt@g#1 )$$}

\def\aligntag#1\endaligntag
  $${\gdef\tag##1\\{&(##1 )\cr}\eqalignno{#1\\}$$
  \gdef\tag##1$${\eqnono(\displayt@g##1 )$$}}

\def\eqalignno#1{\displ@y \tabskip\centering
  \halign to\displaywidth{\hfil$\displaystyle{##}$\tabskip\z@skip
    &$\displaystyle{{}##}$\hfil\tabskip\centering
    &\llap{$\displayt@gpar##$}\tabskip\z@skip\crcr
    #1\crcr}}

\def\displayt@gpar(#1){(\displayt@g#1 )}

\def\displayt@g#1 {\rm\ifunc@lled{#1}\global\advance\tagnumber by1
        {\def\next{#1}\ifx\next\empty\else\expandafter
        \xdef\csname
 @qnnum#1\endcsname{\t@ghead\number\tagnumber}\fi}%
  \writenew@qn{#1}\t@ghead\number\tagnumber\else
        {\edef\next{\t@ghead\number\tagnumber}%
        \expandafter\ifx\csname @qnnum#1\endcsname\next\else
        \w@rnwrite{Equation \noexpand\tag{#1} is
a duplicate number.}\fi}%
  \csname @qnnum#1\endcsname\fi}

\def\ifunc@lled#1{\expandafter\ifx\csname @qnnum#1\endcsname\relax}

\let\@qnend=\end\gdef\end{\if@qnfile
\immediate\write16{Equation numbers
written on []\jobname.EQN.}\fi\@qnend}

\catcode`@=12

\catcode`@=11
\newcount\r@fcount \r@fcount=0
\newcount\r@fcurr
\immediate\newwrite\reffile
\newif\ifr@ffile\r@ffilefalse
\def\w@rnwrite#1{\ifr@ffile\immediate\write\reffile{#1}\fi\message{#1}}

\def\writer@f#1>>{}
\def\referencefile{
  \r@ffiletrue\immediate\openout\reffile=\jobname.ref%
  \def\writer@f##1>>{\ifr@ffile\immediate\write\reffile%
    {\noexpand\refis{##1} = \csname r@fnum##1\endcsname = %
     \expandafter\expandafter\expandafter\strip@t\expandafter%
     \meaning\csname r@ftext
\csname r@fnum##1\endcsname\endcsname}\fi}%
  \def\strip@t##1>>{}}

\def\citeall#1{\xdef#1##1{#1{\noexpand\cite{##1}}}}
\def\cite#1{\each@rg\citer@nge{#1}}	

\def\each@rg#1#2{{\let\thecsname=#1\expandafter\first@rg#2,\end,}}
\def\first@rg#1,{\thecsname{#1}\apply@rg}	
\def\apply@rg#1,{\ifx\end#1\let\next=\relax
\else,\thecsname{#1}\let\next=\apply@rg\fi\next}

\def\citer@nge#1{\citedor@nge#1-\end-}	
\def\citer@ngeat#1\end-{#1}
\def\citedor@nge#1-#2-{\ifx\end#2\r@featspace#1 
  \else\citel@@p{#1}{#2}\citer@ngeat\fi}	
\def\citel@@p#1#2{\ifnum#1>#2{\errmessage{Reference range #1-
#2\space is bad.}%
    \errhelp{If you cite a series of references by the notation M-N, then M
and
    N must be integers, and N must be greater than or equal to M.}}\else%
 {\count0=#1\count1=#2\advance\count1
by1\relax\expandafter\r@fcite\the\count0,
  \loop\advance\count0 by1\relax
    \ifnum\count0<\count1,\expandafter\r@fcite\the\count0,%
  \repeat}\fi}

\def\r@featspace#1#2 {\r@fcite#1#2,}	
\def\r@fcite#1,{\ifuncit@d{#1}
    \newr@f{#1}%
    \expandafter\gdef\csname r@ftext\number\r@fcount\endcsname%
                     {\message{Reference #1 to be supplied.}%
                      \writer@f#1>>#1 to be supplied.\par}%
 \fi%
 \csname r@fnum#1\endcsname}
\def\ifuncit@d#1{\expandafter\ifx\csname r@fnum#1\endcsname\relax}%
\def\newr@f#1{\global\advance\r@fcount by1%
    \expandafter\xdef\csname r@fnum#1\endcsname{\number\r@fcount}}

\let\r@fis=\refis			
\def\refis#1#2#3\par{\ifuncit@d{#1}
   \newr@f{#1}%
   \w@rnwrite{Reference #1=\number\r@fcount\space is not cited up to
 now.}\fi%
  \expandafter
\gdef\csname r@ftext\csname r@fnum#1\endcsname\endcsname%
  {\writer@f#1>>#2#3\par}}

\def\ignoreuncited{
   \def\refis##1##2##3\par{\ifuncit@d{##1}%
    \else\expandafter\gdef
\csname r@ftext\csname r@fnum##1\endcsname\endcsname%
     {\writer@f##1>>##2##3\par}\fi}}

\def\r@ferr{\endreferences\errmessage{I was expecting to see
\noexpand\endreferences before now;  I have inserted it here.}}
\let\r@ferences=\references
\def\references{\r@ferences\def\endmode{\r@ferr\par\endgroup}}

\let\endr@ferences=\endreferences
\def\endreferences{\r@fcurr=0
  {\loop\ifnum\r@fcurr<\r@fcount
    \advance\r@fcurr by
1\relax\expandafter\r@fis\expandafter{\number\r@fcurr}%
    \csname r@ftext\number\r@fcurr\endcsname%
  \repeat}\gdef\r@ferr{}\endr@ferences}


\let\r@fend=\endpaper\gdef\endpaper{\ifr@ffile
\immediate\write16{Cross References written on
[]\jobname.REF.}\fi\r@fend}

\catcode`@=12

\citeall\refto		
\citeall\ref		%
\citeall\Ref		%

\ignoreuncited
\def\hook{\mathbin{\raise2.5pt\hbox{\hbox{{\vbox{\hrule height.4pt
width6pt depth0pt}}}\vrule height3pt width.4pt depth0pt}\,}}
\def\ss{\scriptscriptstyle}

\pageno0
\line{\hfill October 1994}
\title SOME REMARKS ON
GRAVITATIONAL ANALOGS OF MAGNETIC CHARGE
\author C. G. Torre
\affil Department of Physics
Utah State University
Logan, UT 84322-4415
USA
\abstract
\noindent{\bf Abstract:}

Existing mathematical results are applied to the problem of classifying
closed $p$-forms which are locally constructed from Lorentzian metrics
on an
$n$-dimensional orientable manifold $M$ ($0<p<n$).   We show that the
only closed, non-exact forms are generated by representatives of
cohomology classes of $M$ and $(n-1)$-forms representing $n$-
dimensional (with $n$ even)
generalizations of the conservation of ``kink number'', which was exhibited
by Finkelstein and Misner for $n=4$.  The cohomology class that defines
the kink number depends only on the diffeomorphism equivalence class of
the metric, but a result of Gilkey implies that there is no
representative of this cohomology class which is built from the metric,
curvature and covariant derivatives of curvature to any finite order.
\endtitlepage

\noindent{\bf 1. Introduction.}

Let $M$ be a smooth $n$-dimensional manifold and let $g$ be a
Lorentzian
metric on $M$.  Let $\alpha$ be a $p$-form, $0<p<n$, locally constructed
from the metric and its derivatives.  We say that $\alpha$ is a {\it locally
conserved p-form} if $d\alpha=0$ for all choices of $g$.  Here $d$ is the
exterior derivative on $M$.  If $\Sigma$ is a $p$-dimensional closed,
oriented
submanifold of $M$, then, for any
given metric, the integral of $\alpha$ over $\Sigma$ depends only on the
homology class of $\Sigma$.  Wald has called such integrals ``gravitational
analogs of magnetic charge''.  Of course, if there is a $(p-1)$-form $\beta$
that
is locally constructed from the metric and its derivatives such that
$\alpha=d\beta$, then $\alpha$ is trivially closed and the corresponding
charge will vanish for all metrics.   Let us then define a {\it
topological conservation law} \refto{rigidfootnote} as an equivalence class
of locally conserved $p$-forms; two locally conserved $p$-forms, $\alpha$
and $\alpha^\prime$, are
equivalent if there is some $(p-1)$-form $\beta$ that is locally constructed
from the metric and its derivatives such that $\alpha-
\alpha^\prime=d\beta$.

Wald has given a clear and rather complete discussion of topological
conservation
laws in a general field theory \refto{Wald1990a}.  In particular, he has
shown that when
the fields of interest are Lorentzian metrics, the charge of a topological
conservation law can
only depend on the homotopy class of the metric $g$ on $M$.   Then,
using results of Unruh \refto{Unruh1971}, Wald is able to conclude that
for $n=4$ there
are no
topological conservation laws which are covariantly constructed using the
metric, polynomials
in the curvature and covariant derivatives of curvature to any
finite order.  This result is in accord with general results of Gilkey
\refto{Gilkey1975a, Gilkey1975b}.  Gilkey classifies ``natural''
conservation laws, which are closed forms modulo exact forms, all of
which are
constructed covariantly from the metric, curvature and covariant
derivatives of curvature.
Gilkey's work implies that the only natural conservation laws are generated
by the Pontrjagin forms.  These natural
conservation laws only exist when $n>4$.

Quite some time ago, Finkelstein and Misner \refto{Misner1959} studied
homotopy
classes of asymptotically flat Lorentzian metrics on $M=R^4$ and
showed that, given a metric, one can associate an integer to any
asymptotically spacelike
hypersurface in $M$.  This
integer, later dubbed the ``kink number'' \refto{Dunn1991}, represents the
number of
times light cones tumble as one traverses
the hypersurface.  The kink number is a
homotopy invariant of $g$ and is thus unchanged by any continuous
deformation of the hypersurface.  Given a metric, the work of
\refto{Dunn1991} shows
how to evaluate this kink number.  By analogy with similar results from
other
field theories, it is reasonable to suppose that the kink number
corresponds to a topological conservation law, {\it i.e.}, is a
gravitational analog of magnetic charge.  That this is so is, to some extent,
implicit in the
formula for the kink number given in \refto{Dunn1991}.  We would like
to make this fact
explicit as well as provide some additional results on topological
conservation laws for field theories based on Lorentzian metrics on
a general class of $n$-dimensional manifolds.

Our goals in this paper are as follows.

\item{(i)} Give an interpretation of the kink number as a topological
conservation law, that is, as a cohomology class in the Euler-Lagrange
complex \refto{Anderson1992} associated with Lorentzian metrics on any
orientable,
even-dimensional manifold.

\item{(ii)} Show that all topological conservation laws are generated by
cohomology classes of $M$ and the kink conservation law.

\item{(iii)} Show that while the cohomology class that defines the kink
number
conservation law depends only on the diffeomorphism equivalence class of
the metric, there is no naturally
constructed representative of this equivalence class.  In
particular, this result shows how the kink number in dimension four
manages to evade the
rather stringent results of \refto{Wald1990a} and, more generally,
\refto{Gilkey1975a, Gilkey1975b}.

The main technical tool that we shall use is the variational bicomplex
\refto{Anderson1992},
which is specifically tailored to analyze structures such as topological
conservation laws.
We shall describe the results we need from the bicomplex in \S2.  In \S3 we
apply these results to the topological conservation laws built from
Lorentzian
metrics and exhibit a representative of the topological conservation law
corresponding to the kink
number.  We also explain the sense in which this conservation law is
unique.  In \S4 we point out that our construction of the differential form
representing the kink conservation law is not natural, {\it i.e.}, the locally
conserved form is not built from a universal expression in the metric,
curvature, and covariant
derivatives of curvature.  This leaves open the possibility that one can find
another
representative of this conservation law that is naturally constructed.
However, the results of Gilkey imply that there is no
natural representative of this topological conservation law.

\vskip 0.2truein
\taghead{2.}
\noindent{\bf 2. Topological conservation laws and the variational
bicomplex.}

To begin, we need a more precise definition of a topological conservation
law.
For our purposes it is best to give this definition in the context of the
variational bicomplex, although this is certainly not necessary, see for
example \refto{Wald1990a}.  Our treatment is taken from that of
Anderson \refto{Anderson1992, Andersonbook}.

Let $\pi\colon E\to M$ be the bundle of Lorentzian metrics over the
$n$-dimensional manifold $M$.  A
section $g\colon M\to E$ defines a metric on $M$.  Of course we assume
that $M$ admits
global
Lorentzian metrics.  If $M$ is non-compact it always admits a
Lorentzian metric; if $M$ is compact it admits a Lorentzian metric if and
only if the Euler number of $M$ is zero \refto{Marcus1955}.  We denote
by $\pi_{\ss
M}\colon J^\infty(E)\to M$ the infinite jet bundle of metrics.  Here the
bundle is interpreted as having base space $M$. There is also a projection
$\pi_{\ss E}\colon J^\infty(E)\to E$.  A section $g$ of $E$ has a canonical
lift to a section $j^\infty(g)\colon M\to J^\infty(E)$ called the jet of $g$.
For a
general description of jet bundles see \refto{Saunders1989}.

Given local coordinates $x^i$ on $U\subset M$ we have local coordinates
on $J^\infty(\pi^{-1}(U))$ defined by
$$
(x^i, g_{ij}, g_{ij,k}, g_{ij,kl},\ldots),\tag21
$$
where $g_{ij}$ are the components of $g$ in the coordinates $x^i$ and for
any section $g$ of $E$
$$
g_{ij,k_1\cdots k_l}(j^\infty(g))={\partial^lg_{ij}(x)\over\partial
x^{k_1}\cdots\partial x^{k_l}}.
$$

A differential form $\omega$ on $J^\infty(E)$ is called a {\it contact
form} if for every section $g\colon M\to E$,
$$
[j^\infty(g)]^*\omega=0.
$$
The set of contact forms is a differential ideal in the ring
$\Omega^*(J^\infty(E))$ of all differential forms on $J^\infty(E)$.   In the
local coordinates \(21) the contact ideal is spanned locally by the contact
1-forms
$$
\theta_{ij,k_1\cdots k_l}=dg_{ij,k_1\cdots k_l}-g_{ij,k_1\cdots
k_lm}dx^m,\qquad l=0,1,2,\ldots.\tag22
$$
Here, and in all that follows, $d$ is the exterior derivative on
$J^\infty(E)$.

The contact ideal defines a connection on $J^\infty(E)$.  In particular, a
vector $X$ at a point $\sigma \in J^\infty(E)$ is said to be
$\pi_{\ss M}$-vertical if $(\pi_{\ss M})_*X=0$ at $\pi_{\ss M}(\sigma)$;
$X$ is said to be
horizontal at $\sigma$ if $X\hook\omega=0$ for all contact forms
$\omega$ at
$\sigma$.  A
$p$-form $\gamma$ on $J^\infty(E)$ is said to be of type $(r,s)$, where
$r+s=p$, if at each point of $J^\infty(E)$
$$
\gamma(X_1,X_2,\ldots,X_p)=0
$$
whenever more than $s$ of the vectors $X_1,X_2,\ldots,X_p$ are $\pi_{\ss
M}$-vertical, or more than $r$ of the vectors are horizontal.  The space of
type $(r,s)$ forms is denoted $\Omega^{r,s}(J^\infty(E))$.
In the local coordinates \(21) a type $(r,s)$ form is a sum of terms of the
form
$$
f[g]dx^{i_1}\wedge\cdots\wedge dx^{i_r}\wedge\Theta,
$$
where $\Theta$ is a wedge product of $s$ contact forms \(22) and $f[g]$ is
a function on $J^\infty(\pi^{-1}(U))$ depending on the metric
and its derivatives to some finite order.

There is a direct sum decomposition
$$
\Omega^p(J^\infty(E))=\bigoplus_{r+s=p}\Omega^{r,s}(J^\infty(E)),
$$
and we let
$\pi^{r,s}\colon\Omega^p(J^\infty(E))\to\Omega^{r,s}(J^\infty(E))$
denote the projection to $\Omega^{r,s}(J^\infty(E))$, where $p=r+s$.  The
exterior derivative on $J^\infty(E)$,
$$
d\colon\Omega^p(J^\infty(E))\to\Omega^{p+1}(J^\infty(E)),
$$
splits into a horizontal and vertical piece via
$$
d=d_{\ss H}+d_{\ss V},
$$
$$
d_{\ss H}\colon\Omega^{r,s}(J^\infty(E))\to\Omega^{r+1,s}(J^\infty(E))
$$
$$
d_{\ss V}\colon\Omega^{r,s}(J^\infty(E))\to\Omega^{r,s+1}(J^\infty(E))
$$
where, for a $p$-form $\gamma\in\Omega^{r,s}(J^\infty(E))$,
$$
d_{\ss H}\gamma=\pi^{r+1,s}(d\gamma)
$$
and
$$
d_{\ss V}\gamma=\pi^{r,s+1}(d\gamma).
$$

As an example, in local coordinates
the horizontal exterior derivative of a function
$f\colon J^\infty(E)\to R$ takes the form
$$
d_{\ss H}f = ({\partial f\over\partial x^i} + {\partial f\over\partial
g_{ij}}g_{ij,m}+{\partial f\over\partial g_{ij,k}}g_{ij,km}+\cdots)dx^m
=(D_mf)dx^m,\tag26
$$
and the vertical exterior derivative of $f$ takes the form
$$
d_{\ss V}f={\partial f\over\partial g_{ij}}\theta_{ij}
+{\partial f\over\partial g_{ij,k}}\theta_{ij,k}+\cdots\ .
$$
In particular, because
$$
dg_{ij}=\theta_{ij}+g_{ij,k}dx^k,\tag23
$$
we have that
$$
d_{\ss V}g_{ij}=\theta_{ij},\tag24
$$
and
$$
d_{\ss H} g_{ij}=g_{ij,k}dx^k = (D_k g_{ij})dx^k.\tag25
$$

The differential operator $D_i$ in \(26) and \(25) is the {\it total
derivative} operator.
It is not too hard to see that the horizontal exterior derivative corresponds
to the usual exterior derivative on $M$, with partial differentiation
replaced by total differentiation.  More precisely, if $g\colon M\to E$ is a
metric, $\alpha\in\Omega^{r,0}(J^\infty(E))$, and $d_{\ss M}$ is the
exterior
derivative on $M$, then we have,
$$
j^\infty(g)^*(d_{\ss H}\alpha)=d_{\ss
M}[j^\infty(g)^*(\alpha)].
$$
In this way the notion of ``exterior derivative of a form locally constructed
from the metric'' is made precise in terms of $d_{\ss H}$ acting on forms
in $\Omega^{r,0}(J^\infty(E))$.

The identity $d^2=0$ now decomposes into
$$
d_{\ss H}^2=0,\qquad d_{\ss V}^2=0,\qquad d_{\ss H}d_{\ss V}
=-d_{\ss V}d_{\ss H},
$$
so that the de Rham complex $\Omega^*(J^\infty(E))$ on the jet bundle
decomposes into a double complex called the {\it variational bicomplex}.
Topological conservation laws arise as cohomology classes on the ``bottom
edge'' of the bicomplex.  More precisely, a {\it locally conserved p-form}
is a form $\alpha\in\Omega^{p,0}(J^\infty(E))$ such that
$$
d_{\ss H}\alpha=0.\tag27
$$
If $\alpha$ is exact, then there is a form $\beta\in\Omega^{p-
1,0}(J^\infty(E))$ such
that
$$
\alpha=d_{\ss H}\beta.\tag28
$$
The restriction of the de Rham complex on $J^\infty(E)$ to the forms in
$\Omega^{p,0}(J^\infty(E))$, $0<p<n$, will be called the
{\it Euler-Lagrange complex} \refto{footnoteEuler}, and
will be denoted ${\cal E}^*(J^\infty(E))$.  Cohomology classes in ${\cal
E}^*(J^\infty(E))$ are forms which are $d_{\ss H}$-closed \(27) modulo
forms that are
$d_{\ss H}$-exact \(28).  Thus a {\it topological conservation law} is a
cohomology
class in ${\cal E}^*(J^\infty(E))$.

We now present two results from the theory of the variational bicomplex
that vastly simplify the computation of all topological conservation laws.
Let
$H^p(\Omega^*(J^\infty(E)))$ denote the $p^{th}$ cohomology of
the de Rham complex on $J^\infty(E)$.  Similarly, denote by $H^p({\cal
E}^*(J^\infty(E)))$ the $p^{th}$ cohomology of the Euler-Lagrange
complex.  It can be shown \refto{Anderson1992, Andersonbook} that there
is an isomorphism between these
vector spaces.   More precisely, define a map
$\Psi\colon\Omega^p(J^\infty(E))\to{\cal E}^p(J^\infty(E))$ for $0<p<n$
by
$$
\Psi(\alpha)=\pi^{p,0}(\alpha).\tag29
$$
The induced map
$$
\Psi^*\colon H^p(\Omega^*(J^\infty(E)))\to H^p({\cal E}^*(J^\infty(E)))
\tag210
$$
is an isomorphism.  Next, it can be shown that the projection $\pi_{\ss
E}\colon J^\infty(E)\to E$ is a homotopy equivalence, and hence the de
Rham cohomology of $J^\infty(E)$ is isomorphic to the de Rham
cohomology of $E$.  Thus {\it the topological conservation
laws are in one to one correspondence with the cohomology classes of E}.
In detail, the correspondence just stated is as follows.  Let $\alpha$ be a
closed $p$-form representing a nontrivial cohomology class in
$\Omega^p(E)$.  The form $\alpha$ can be pulled back via $\pi_{\ss E}$
to give a representative of a nontrivial cohomology class in
$\Omega^p(J^\infty(E))$.  The map $\Psi$ in \(29) then defines a
representative of a nontrivial cohomology class in ${\cal
E}^p(J^\infty(E))$.  The theory of the variational bicomplex tells us that
all topological conservation laws arise in this manner.

\vskip 0.2truein
\taghead{3.}
\noindent{\bf 3.  Classification of topological conservation laws.}

The theory of the variational bicomplex reduces the task of computing the
cohomology classes of the Euler-Lagrange complex to that of the bundle of
metrics.  According to Steenrod \refto{Steenrod1951}, there is a
deformation retraction
$\varphi\colon E\to E^\prime$ of the bundle of metrics $E$ to a bundle
$\pi^\prime\colon E^\prime\to M$ which has the same bundle data as $E$
except
the fiber $F^\prime$ is the real projective space $RP^{n-1}$.  This
deformation retraction corresponds to the construction of a
line element field from a given Lorentzian metric and an auxiliary
Riemannian metric \refto{Hawking1973}.  Thus we have the isomorphism
$$
H^*(E)=H^*(E^\prime).\tag30
$$

To begin, let us assume that $M$ is parallelizable, in which case the bundle
of metrics is trivial, that is, $E=M\times F$, and hence
$
E^\prime=M\times F^\prime.
$
The Kunneth formula \refto{Bott1982} then shows that the cohomology of
$E^\prime$ is given by
$$
H^*(E^\prime)=H^*(M)\otimes H^*(F^\prime).\tag212
$$
The cohomology classes of $M$ are smooth
invariants of $M$.  The closed forms on $E$ representing this cohomology
can be obtained by pulling back representatives on $M$ using $\pi$.  These
locally conserved forms are manifestly independent of the metric.
The cohomology of $F^\prime$, {\it i.e.}, $H^p(RP^{n-1})$, $p>0$, is
trivial if $n$ is odd, and when $n$ is even the only nontrivial class is
at form degree $n-1$:
$$
H^{n-1}(RP^{n-1})=R, \quad n\ {\rm even}.\tag211
$$
A representative of $H^{n-1}(F^\prime)$ can be obtained by taking, {\it
e.g.}, the standard volume form on the $(n-1)$-sphere, which projects to
give a volume form on $RP^{n-1}$ (when $n$ is even) via the usual
antipodal projection from
$S^{n-1}$ to $RP^{n-1}$.  According to \(212), this form on $RP^{n-1}$
corresponds to a representative of a topological conservation law.  A
formula
for this locally conserved form can be constructed as follows.   Keeping in
mind that we are assuming for the moment that $M$ is parallelizable, fix a
global
trivialization $e\colon E^\prime\to M\times F^\prime$.  Let $\pi_{\ss
F^\prime}\colon M\times F^\prime
\to F^\prime$ denote the projection to the fiber defined by the
trivialization.  Assuming $n$ is even,
(i)
Take a volume form $\Omega$ on $F^\prime=RP^{n-1}$ and pull it back
to $E^\prime$ via
$\pi_{\ss F^\prime}$.   Using
$\varphi$, pull the resulting form back to $M\times F$.   The resulting
form is then
pulled back to $J^\infty(M\times F)$ using the projection $\pi_{\ss E}$.
Finally,
apply the map $\Psi$ to construct the representative $\alpha$ of the
topological
conservation law:
$$
\alpha = \pi^{n-1,0}\{\pi^*_{\ss E}[\varphi^*(\pi^*_{\ss
F^\prime}\Omega)]\}
= \pi^{n-1,0}\{(\pi_{\ss F^\prime}\circ\varphi\circ\pi_{\ss
E})^*\Omega\}.\tag213
$$

Having constructed the closed $(n-1)$-form $\alpha$, we can compute the
charge $Q[g]$ associated with a metric as follows.  Given a metric
$g\colon M\to E$ with jet $j^\infty(g)\colon M\to J^\infty(E)$, we pull
$\alpha$ back to give an $(n-1)$-form on $M$ which we integrate over a
closed, oriented $(n-1)$-dimensional submanifold $\Sigma$
$$
Q[g]=\int_\Sigma [j^{\infty}(g)]^*\alpha.\tag214
$$
Up to a numerical factor, this
integral represents the degree of the map from $\Sigma$ to $RP^{n-1}$
defined by the metric.  When $n=4$ and $\Sigma=S^3$, the integral \(214)
corresponds to the formula proposed in \refto{Dunn1991} for the kink
number.  Accordingly, we shall call the conservation law represented by
$\alpha$ the {\it kink conservation law}.

When $M$ is not parallelizable, {\it i.e.}, when $E^\prime$ is not trivial
we can
repeat the construction above using a {\it local} trivialization, in which
$\pi^{\prime-1}(U)=U\times F^\prime$.  However, there may
be obstructions to patching together the local representative \(213) of the
cohomology class on $U\times F^\prime$ to give a global representative of
a class on $E^\prime$.  In
order to generalize the kink conservation law to non-parallelizable
manifolds we should find a closed $(n-1)$-form on $E^\prime$ whose
restriction to any fiber generates the cohomology \(211) of the fiber.
Moreover, if we can do this, then \(30) and the Leray-Hirsch theorem
\refto{Bott1982} imply that
we still have the isomorphism
$$
H^*(E)=H^*(M)\otimes H^*(F^\prime).\tag2141
$$
Because $F^\prime$ is cohomologically trivial when $n$ is odd, it
immediately follows that the Leray-Hirsch theorem is applicable in this
case, and we have that
$$
H^p(E)=H^p(M), \qquad n\ {\rm odd}.
$$
When $n$ is even, the obstructions to constructing a global $(n-1)$-form
on $E^\prime$ which restricts to generate the cohomology \(211) are (i)
orientability of the bundle $E^\prime$ and (ii) the Euler class of
$E^\prime$.  This result follows directly from the discussion of sphere
bundles in \refto{Bott1982}, which is easily adapted to the case where the
sphere is
replaced by $RP^{n-1}$.  In essence, the results of \refto{Bott1982} are
unchanged
because $H^*(S^{n-1})=H^*(RP^{n-1})$ when $n$ is even.  The Euler
class of $E^\prime$ vanishes because we assume that $E$ admits a global
section, that is, a global Lorentzian metric, which implies that $E^\prime$
admits a global section.  Furthermore, because the transition functions of
$E^\prime$ are induced by those of the tangent bundle, $E^\prime$ is
orientable if and only if $M$ is orientable.  Hence, if $M$ is orientable, a
representative of the kink conservation law can be defined globally, and the
isomorphism \(2141) holds.

Thus, aside from cohomology classes of
$M$, the kink number arises as an additional topological conservation law
when $n$ is even
and $M$ is orientable.  All topological conservation laws for Lorentzian
metrics on
orientable manifolds are generated by these conservation laws.

\vskip 0.2truein
\taghead{4.}
\noindent{\bf 4. General covariance of the conservation law.}

Our construction \(213) of a (local) representative $\alpha$ of the kink
conservation
law depends on the deformation retraction $\varphi$ and the choice of
(local) trivialization.  This dependence means that $\alpha$ cannot be
``naturally'' constructed from the spacetime metric and its derivatives.
In order to make this discussion more precise, we phrase
it in terms of the behavior of $\alpha$ with respect to the action of
spacetime diffeomorphisms on the metric \refto{Atiyah1973}.

Let
$\Psi\colon M\to M$ be a diffeomorphism.  The map $\Psi$ lifts to give a
bundle map $\Psi_{\ss E}\colon E\to E$ via $(x,g)\longrightarrow
(\Psi(x),\Psi_*g)$.  The bundle map $\Psi_{\ss E}$, in turn, lifts by
prolongation \refto{Saunders1989} to give a bundle map ${\rm
pr\,}\Psi\colon J^\infty(E)\to
J^\infty(E)$.  Let us consider a $p$-form $\rho$ locally constructed from
the metric and its derivatives.  Such a form is a map from $J^\infty(E)$
into the bundle of $p$-forms on $M$, $\rho\colon J^\infty(E)\to
\Omega^p(M)$.  We say that $\rho$ is a {\it natural p-form} if for every
point $\sigma\in J^\infty(E)$ and for every diffeomorphism $\Psi$
$$
\rho({\rm pr\,}\Psi(\sigma))=(\Psi_*\rho)(\sigma).
$$
The notion of a natural $p$-form gives a precise characterization of a
``$p$-form covariantly constructed from the metric and its derivatives''.
Of course, the property of being natural can be generalized to any type of
tensor field.  Natural tensor fields are defined on any manifold by
universal formulas involving the metric and its derivatives.  It is an old
result of Thomas that natural tensor fields are
locally constructed from the metric, the curvature,
and covariant derivatives of the curvature to some order
\refto{Thomas1934}.  If $M$ is
oriented, then we should restrict attention to orientation preserving
diffeomorphisms.  In this case, natural tensor fields are constructed in a
tensorial
fashion from the metric, curvature, covariant derivatives of curvature, and
the volume form defined by the metric.

It is straightforward to verify that $\alpha$ in \(213) is not a natural
$(n-1)$-form.  For
example,
in local coordinates \(21) on $J^\infty(\pi^{-1}(U))$ the components of
$\alpha$
are locally constructed
from the metric and its first derivatives only (this follows from \(23)--
\(25)).
Thomas's result then implies that $\alpha$ is not a natural form.   In the
bundle language we have been using, we say that $\alpha$ does not
behave naturally under the map ${\rm pr\,}\Psi$.  Thus, while $\alpha$ is
globally defined on any orientable even-dimensional manifold, its
definition depends on the choice of manifold.  On the
other hand, given the isomorphism \(210), the bundle diffeomorphism
${\rm
pr\,}\Psi$ cannot change the
cohomology class of $\alpha$.  Thus, we conclude that while $\alpha$ will
change unnaturally under the action of a diffeomorphism, it can only do so
by the addition of a $d_{\ss H}$-exact $(n-1)$-form locally constructed
from the metric and its derivatives:
$$
\alpha({\rm pr\,}\Psi(\sigma))-(\Psi_*\alpha)(\sigma)=d_{\ss
H}\tau(\sigma).$$
In other words, the charge $Q$ in \(214)
{\it is}
diffeomorphism invariant even if its integrand is not.

The diffeomorphism invariance of the equivalence class of $\alpha$
suggests that it may be possible to find a natural representative of this
conservation law.  In other words, is there a form $\beta\in\Omega^{n-
2,0}(J^\infty(E))$ such that
$$
\alpha^\prime=\alpha + d_{\ss H}\beta
$$
is a natural $(n-1)$-form with respect to orientation preserving
diffeomorphisms?  To answer this question we observe that such a
natural
$d_{\ss H}$-closed $p$-form $\alpha^\prime\in\Omega^{p,0}(J^\infty(E))$
falls
under the scope of the theorem of Gilkey \refto{Gilkey1975a,
Gilkey1975b}.  Gilkey's theorem serves
to classify
natural $d_{\ss H}$-closed $p$-forms $\rho$ modulo forms $d_{\ss
H}\gamma$
where $\gamma$ is a natural $(p-1)$-form.  In this setting of {\it
equivariant
cohomology}, Gilkey asserts that if $\rho$ is a natural $d_{\ss H}$-closed
$p$-form, then there is a natural $(p-1)$-form $\gamma$ such that
$$
\rho=\kappa + d_{\ss H}\gamma,
$$
where $\kappa$ is a {\it characteristic form}, {\it
i.e.}, an
element of the
algebra generated by the Pontrjagin forms \refto{footnote_equivariant}.
This means that all natural forms which are $d_{\ss H}$-closed but are not
$d_{\ss H}$ of a natural form are of even degree $4k$, where
$k=1,2,\ldots$.  We
have seen that the kink
conservation law only exists in odd degree.  Hence there is no natural
representative of the kink conservation law.

The results of this paper are independent of any field equations that might
be imposed on the metric, {\it e.g}, the vacuum Einstein equations.  In the
presence of field equations, new conservation laws may be obtained since
the differential forms need only be closed, say, when the Einstein tensor
vanishes.  The variational bicomplex for Lorentzian metrics can be pulled
back to the jet bundle of Einstein metrics and the conservation laws can be
classified using spinor methods \refto{CGT1993, cmp}.  The results of this
investigation will be presented elsewhere \refto{bicomplex}.
\vskip 0.25truein
\noindent{\bf Acknowledgments}

I would like to thank Chris Isham, Jim Stasheff, and especially Ian
Anderson for very helpful correspondence.

\references

\refis{cmp}{I. M. Anderson and C. G. Torre, ``Classification of
Generalized Symmetries for the Vacuum Einstein Equations'', Utah
State
University preprint, (1994).}

\refis{rigidfootnote}{Topological conservation laws, as defined here,
are a
subset of what are known as ``rigid conservation laws''; see
\refto{Anderson1992}.}

\refis{Andersonbook}{I. M. Anderson, {\it The Variational Bicomplex},
book to appear.}

\refis{footnoteEuler}{The true Euler-Lagrange complex is actually an
extension of the complex being used here; see \refto{Anderson1992}.
We
suppress the extension as it plays no role in this paper.}

\refis{footnote_equivariant}{Gilkey's results, and generalizations of
these
results, can be understood in terms of the equivariant cohomology
of the
variational bicomplex for (pseudo-) Riemannian structures
\refto{Anderson1992}.}

\refis{bicomplex}{I. M. Anderson and C. G. Torre, ``Variational
Bicomplexes Associated to the Einstein Equations'', in
preparation.}

\refis{Wald1990a}{R. Wald, \jmp 31, 2378, 1990.}

\refis{Unruh1971}{W. G. Unruh, \grg 2, 27, 1971}.

\refis{Gilkey1975a}{P. Gilkey, \journal Math. Scand., 36, 109,
1975}.

\refis{Gilkey1975b}{P. Gilkey, \journal Ann. Math., 102, 187,
1975}.

\refis{Misner1959}{D. Finkelstein and C. Misner, \ann 6, 320,
1959.}

\refis{Dunn1991}{K. Dunn, T. Harriott, and J. Williams, \jmp
32, 476,
1991.}

\refis{Anderson1992}{I. M. Anderson, ``Introduction to the
Variational
Bicomplex'', in {\it Mathematical Aspects of Classical Field
Theory} (Eds.
M. Gotay, J. Marsden, V. Moncrief),\journal Cont. Math., 32,
51, 1992.}

\refis{Marcus1955}{L. Markus, \journal Ann. Math., 62, 411,
1955.}

\refis{Saunders1989}{D. Saunders, {\it The Geometry of Jet
Bundles},
(Cambridge University Press, Cambridge 1989).}

\refis{Steenrod1951}{N. Steenrod, {\it The Topology of Fibre
Bundles},
(Princeton University Press, Princeton 1951)}.

\refis{Hawking1973}{S. Hawking and G. Ellis, {\it The Large
Scale
Structure of Space-Time}, (Cambridge University Press,
Cambridge
1973).}

\refis{Bott1982}{R. Bott and L. Tu, {\it Differential Forms in
Algebraic
Topology}, (Springer-Verlag, New York 1982).}

\refis{Atiyah1973}{M. Atiyah, R. Bott, and V. Patodi, \journal
Inv. Math.,
19, 279, 1973.}

\refis{Thomas1934}{T. Y. Thomas, {\it Differential Invariants
of
Generalized Spaces}, (Cambridge University Press, Cambridge
1934).}

\refis{CGT1993}{C. G. Torre  and I. M. Anderson, \prl 70,
3525, 1993.}

\endreferences
\bye